# Stark-modulated Rydberg dissipative time crystals at room-temperature applied to sub-kHz electric-field sensing

Darmindra Arumugam

Jet Propulsion Laboratory, California Institute of Technology, Pasadena, 91109, California, USA

E-mail: darmindra.d.arumugam@jpl.nasa.gov;

**Abstract**
Out-of-equilibrium Rydberg gases exhibit emergent many-body phases due to mode competition. Sustained limit cycle oscillations (OSC) emerge when driven by B-fields at room-temperature, forming robust Rydberg dissipative time crystals (DTC). These driven-dissipative Rydberg DTC have recently been shown to develop an effective transition centered at the OSC frequency (-10dB bandwidth of ~1.7kHz, centered at 9.8kHz). Weak RF signals injected within this emergent transition perturb and emerge on the OSC spectrum, from which sensitive and high-resolution sensing of E-fields (~1.6-2.3 $\mu Vcm^{-1}Hz^{-1/2}$) near the OSC frequencies can be achieved. In this article, it is demonstrated that DC and AC Stark fields in the sub-kHz regime can be used effectively to shift (DC) or modulate (AC) the OSC frequency of Rydberg DTC at room-temperature. The AC-Stark driven modulation of the OSC is shown as an effective technique to sense weak AC E-fields in the sub-kHz regime. With a modest setup, a sensitivity of ~7.8 $\mu Vcm^{-1}Hz^{-1/2}$ for AC signals at 300Hz (~8.7× improvement over state-of-art Rydberg atom techniques), and high-resolution detection to as low as sub-Hz is demonstrated. This approach enables the development of ultra-compact ($\ll \lambda/10^6$), extremely low-frequency E-field detectors for applications in remote sensing, communications, navigation, and bio-medical technologies.

Driving and dissipation in quantum systems can give rise to many-body phases[1-3]. In ensembles of Rydberg atoms out of equilibrium, mode-competition between nearby states permits the study of many-body interactions[4,5]. Sustained limit cycle oscillations[4,5] (OSC) emerge when driven by magnetic-fields (B-fields) at room temperature[6] due to the nonlinear energy shifts and competition between distinct Rydberg sublevels, giving rise to long-range time crystalline order[7-15] which manifests as a robust dissipative time crystal (DTC)[6] – characterized by spontaneously broken time-translation symmetry, long-range temporal order, robustness against perturbations, and collective many-body interactions. Supplementary Section 1 addresses key features of DTCs with Rydberg atoms as observed here and in prior works.

Recent research[16] has shown that Rydberg DTCs at room-temperature can develop an emergent and effective transition centered at the OSC frequency which can support very-low-frequency (VLF) electric-field (E-field) sensing (-10dB bandwidth of ~1.7kHz, centered at 9.8kHz). Weak RF signals injected within this emergent transition perturb and emerge on the OSC spectrum, from which sensitive and high-resolution sensing of E-fields (~1.6-2.3 $\mu Vcm^{-1}Hz^{-1/2}$) near the OSC frequencies can be achieved. As the mode-competition are directly due to nearby sublevels, shifts of those sublevels via DC/AC Stark shifting are expected to drive shifts in the OSC frequencies. This work demonstrates that DC and AC Stark fields in the sub-kHz regime can effectively shift (DC) or modulate (AC) the OSC frequency of Rydberg dissipative time crystals at room temperature. The AC Stark-driven modulation enables precise sensing of weak AC E-fields in the sub-kHz range. Here, the modulation appears as a frequency modulation (FM) of the OSC frequency and is easily extracted via standard demodulation techniques by detecting the tone at the peak of the modulation sideband. With a modest setup, a sensitivity of ~7.8 $\mu Vcm^{-1}Hz^{-1/2}$ at 300Hz — an ~8.7× improvement over state-of-the-art Rydberg techniques[17] is achieved, along with high-resolution detection down to sub-Hz levels. The long wavelengths (λ>300km) at sub-kHz frequencies limit the efficiency, sensitivity, and bandwidth of classical compact electrically small ($\ll \lambda$) antennas due to Chu's limit[18]. The method presented in this article paves the way for ultra-compact ($\ll\lambda/10^6$ for sub-kHz RF detectors) and extremely low-frequency (ELF) E-field detectors, with significant and impactful applications where compact electrically small detectors are needed in remote sensing, communications, navigation, and biomedical technologies.

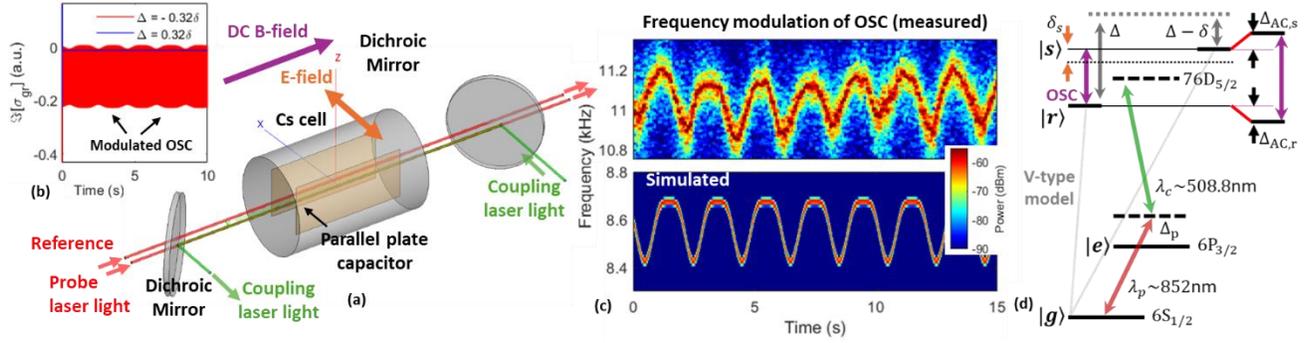

**Fig. 1:** (a) Experimental setup, simulated Stark modulations, energy diagram, and spectrogram measurements. (a) Probe and coupler laser light is counter-propagated in a room-temperature vapor cell containing Cesium atoms. A parallel plate capacitor is embedded in the cell to drive low-frequency electric fields in the sub-kHz regime, and a DC magnetic field (B-field) is used to drive mode-competition at the excited Rydberg states to give limit cycle oscillations (OSC). A reference probe light is used for classical noise reduction. (b) Time-domain simulations of mode-competition between two excited states using a V-type atomic structure, driven additionally by a Stark modulation field at 0.5Hz (2 second period) is shown within (red) and outside (blue) of the OSC regime. In the OSC regime (red), the field modulates the energy levels of the two excited states and results in a modulation of the OSC frequency in the form of a frequency modulation. (c) Measurements and simulations of OSC at room-temperature is shown for a Stark field at 0.5Hz, 3.07mV/cm in a spectrogram. The low frequency of the Stark field (0.5Hz) slowly modulates the OSC frequency and is seen in the spectrogram as an approximately sinusoidal variation at 0.5Hz. Simulation based on a V-type Stark modulated atom system gives the same behavior. (d) The energy diagram in the ladder configuration used for experiments are composed of a probe laser driving to the first excited state at $6P_{3/2}$ and is detuned by $\Delta_p/2\pi$ = 22MHz, a coupler driving to $76D_{5/2}$ and is detuned by $\Delta_c/2\pi$ = 30MHz, and a B-field of 4G is used. The Stark field modulates time-dependent shifts in the excited states that perturb the energy differences and OSC frequency. Simulations use a V-type model (gray line).

## Results

### Observed OSC Stark shifts and modulation

Atomic sensors use coherent quantum systems for ultra-sensitive detection[19]. Rydberg (Cs/Rb) atoms, respond efficiently to microwave and millimeter-wave fields[20-22]. Electromagnetically induced transparency (EIT)[23] enables detection via laser-driven transmission changes. However, highly sensitive sub-kHz-frequency E-field detection is challenging and impractical due to low photon energy, typically requiring off-resonant or indirect methods[17]. AC Stark modulation of room-temperature Rydberg DTCs provides a platform to improve sensitivity of sub-kHz E-field detection.

A probe (~852nm) and coupler (~509nm) laser are counter-propagated in a Cs (Cesium-133) vapor cell (see Fig. 1a) at room-temperature. The probe laser drives to the first excited state at $6P_{3/2}$ and is detuned by $\Delta_p/2\pi$ = 22MHz, a coupler driving to $76D_{5/2}$ and is detuned by $\Delta_c/2\pi$ = 30MHz. A B-field of 4G is turned on, giving rise to splitting to sublevels and mode-competition, that results in a OSC[6,16] observed in the probe laser transmission (see simulation in Fig. 1b, red). The OSC characteristics and determination of B-field with the present setup is given in a recent article[16]. When driven by an ELF AC Stark E-field at $f_{AC}$ =0.5Hz, the OSC time-domain response is modulated at $f_{AC}$ (Fig. 1b, red, simulated). Fig. 1c shows experiment (top) and simulated (bottom) results for $E_{AC}$ =3.07mV/cm in a spectrogram (time-frequency). The Stark field, with $f_{AC}$

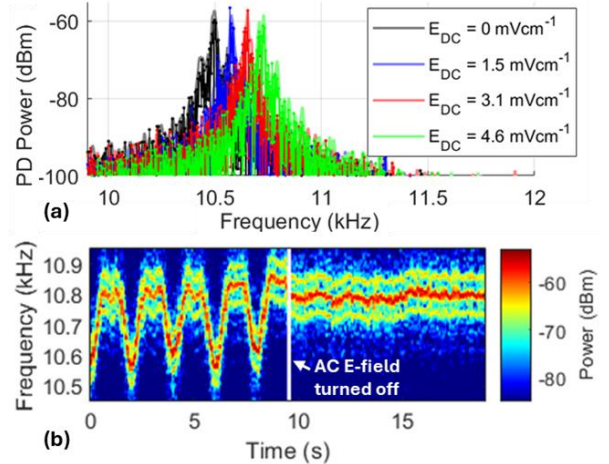

**Fig. 2:** (a) Limit-cycle oscillations (LCO) at n = 76 and B = 4G under different DC Stark-field (no modulation in Stark field) magnitudes ($E_{DC}$ up to 4.6mV/cm). The OSC frequency shifts gradually from 10.49kHz to 10.73kHz due to DC Stark shifting of the Rydberg states where mode-competition, LCO, and OSC arises. (b) A strong AC Stark modulation field ($E_{AC}$) is turned on with $f_{AC}$ =0.5Hz for 9.56s and then turned off. Stark-driven modulation of the OSC is seen at 0.5Hz. When $E_{AC}$ is turned off, the OSC frequency reverts to steady state at ~10.8kHz and is no longer AC modulated.

=0.5Hz, modulates the OSC frequency, appearing as an approximately sinusoidal variation with a periodicity at 0.5Hz. The energy diagram is given in Fig. 1d. The B-field drives sublevels $|s\rangle$ and $|r\rangle$ into mode-competition and an OSC emerges. Shifts in the sublevel energies due to Stark modulation dynamically perturb the OSC frequencies. Spectrogram simulations using a V-type atom (gray, Fig. 1d) reproduce experiments closely in Fig. 1c (bottom, simulated). Fig. 2a shows the steady-state OSC spectrum from the probe transmission (via balanced photodetection, PD power) due to a DC E-field up to $E_{DC}$ =4.6mV/cm. The OSC frequency varies between 10.49kHz ($E_{DC}$ =0V/cm) to about 10.72kHz, showing a direct correlation between DC field amplitude and OSC frequency shift. In Fig. 2b, a strong AC Stark modulation field ($E_{AC}$) is applied at $f_{AC}$ =0.5Hz for ~9.5s before being switched off. During this period, Stark-driven modulation of the OSC is observed at 0.5Hz. Once $E_{AC}$ is turned off, the OSC frequency stabilizes at approximately 10.8kHz, returning to a steady state without AC modulation. Details of the experimental setup and systems are given in Methods and Supplementary Section 2.

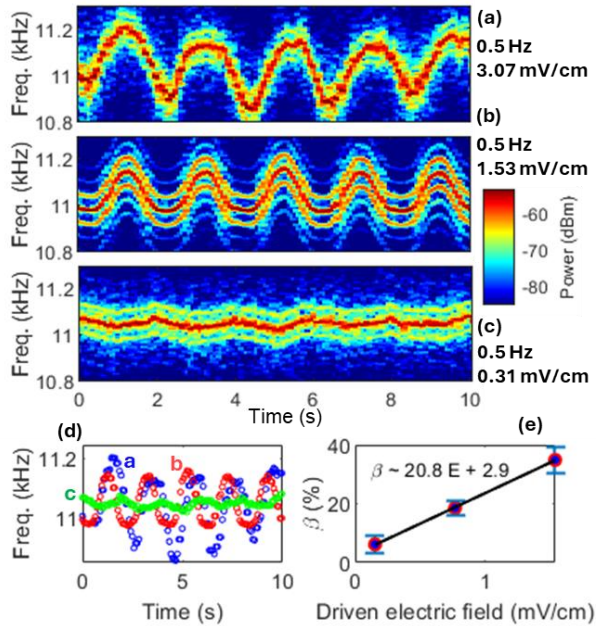

**Fig. 3: Spectrograms and analysis of probe transmission signal to study OSC frequency deviation as a function of AC field magnitude.** (a) AC Stark modulation frequency $f_{AC}$ =0.5Hz and $E_{AC}$ =3.07mV/cm. (b) $E_{AC}$ =1.53mV/cm. (c) $E_{AC}$ =0.31mV/cm. (d) Extracted peak points of each spectrogram dataset is plotted for each, showing gradual reduction in frequency deviation. (e) Estimation of modulation depth, β, as a function of $E_{AC}$. Modulation depth is determined as the ratio of peak deviation to $f_{AC}$ (Δf/ $f_{AC}$) in percentages. A linear fit (black line), linear fit equation, and error bar is given.

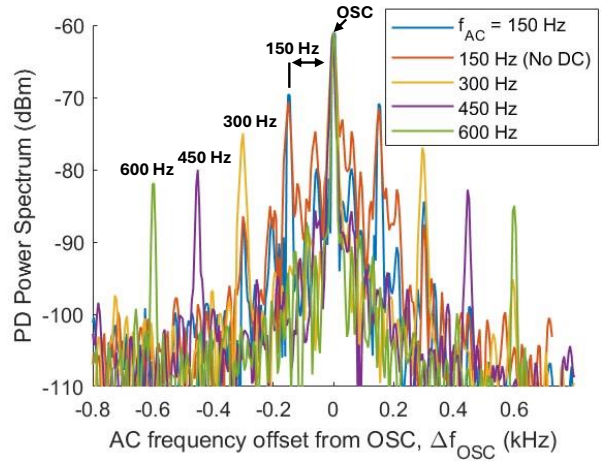

**Fig. 4: Measurements of probe spectrum with OSC under different and higher AC stark modulation frequencies $f_{AC}$, and with and without a DC offset at $f_{AC}$ =150Hz (blue vs red) to show differences in modulation peak.** The spectral measurements (as opposed to spectrogram) show frequency modulation (FM) sidebands at ±$f_{AC}$ on either side of the OSC frequency. To simplify comparisons, each are plotted as a function of (horizontal axis) Δ$f_{OSC}$ (OSC shifted and centered at 0Hz). $E_{AC}$ =1.54mV/cm in all cases. At higher modulation frequencies (e.g.: >300Hz), peak at $f_{AC}$ gradually reduces in intensity, however the noise base is also reduced from being further away from the skirts of the LCO's OSC. The DC field of $E_{DC}$ =19.35mV/cm slightly increases the modulation sideband peak at $f_{AC}$ =150Hz.

**Dependence on E-field magnitude and frequency**

To study AC Stark E-field magnitude ($E_{AC}$) dependence, first the spectrogram measurement of the probe transmission spectrum is studied at $f_{AC}$ =0.5Hz for $E_{AC}$ =3.07, 1.53, 0.31mV/cm in Fig. 3a-c (top to bottom). Due to the strong field and low-frequencies, the frequency deviation (peak-to-peak) is seen to reduce as a function of reduced E-field magnitude. The OSC frequency as a function of time is extracted and shown in Fig. 3d, and the estimation of modulation depth, β, as a function of $E_{AC}$ is given in Fig. 3e. The modulation depth, β, is determined as the ratio of peak deviation to $f_{AC}$ (Δf/ $f_{AC}$) in percentages. The observed linear (see inset of Fig. 3e) relationship between the inferred modulation depth ($\beta$) and the $E_{AC}$ suggests that the system responds linearly to the electric field magnitude. This linearity enables the system to function as a sensitive detector for measuring E-fields. The AC Stark effect induces periodic shifts in the atomic energy levels, leading to a corresponding modulation of the OSC frequency, characteristic of frequency modulation (FM). Since FM signals encode information in frequency variations via their sidebands, the modulation can be spectrally isolated, allowing the frequency of the modulated component, $f_{AC}$, to be extracted with high selectivity. Fig. 4 studies the probe spectrum of OSC under various AC E-field frequencies ($f_{AC}$ = 150, 300, 450, 600Hz). For easier comparison, all data are shown as a function of $\Delta f_{OSC}$ (OSC shifted and centered at 0 Hz), with $E_{AC}$ =1.54 mV/cm in all cases. The spectrum shows sidebands at $\pm f_{AC}$ of the OSC frequency. At higher modulation frequencies (e.g., >300 Hz), the peak at $f_{AC}$ gradually decreases in intensity, while the noise floor also reduces as it moves further from the skirts of the OSC. For the 150Hz measurements, a DC E-field is added and optimized to $E_{DC}$ =19.35mV/cm as an attempt to favorably sensitize the atoms and maximize the signal-to-noise (SNR) (red, Fig. 4). The DC Stark field slightly enhances the modulation sideband peak.

### Sensitivity to AC E-fields and under added DC fields

To study the sensitivity of the technique, the AC Stark E-field magnitude is varied between 1.535mV/cm to 15.35μV/cm for $f_{AC}$ at 150Hz and 300Hz with a measurement resolution bandwidth of 10Hz. Fig. 5a shows a sensitivity analysis. At 150Hz, measurements with and without a DC Stark field ($E_{DC}$ = 19.35mV/cm, near the optimal point) give an $E_{AC}$ field sensitivity of 20.79μVcm$^{-1}$Hz$^{-1/2}$ and 26.5μVcm$^{-1}$Hz$^{-1/2}$, with the optimized DC field improving sensitivity by about ~27%. At 300Hz without DC field, spectral noise reduction improves sensitivity to 7.88μVcm$^{-1}$Hz$^{-1/2}$. Curve fits (black lines) show a log-log fit of measurements, and an analysis of deviation of the measured data from the fits are provided in Supplementary Section 3. To study dependence of SNR to DC field, $E_{DC}$ is varied up to 24.57mV/cm for $f_{AC}$ =150Hz in Fig. 5b. The estimated SNR in dB is shown as a function of $E_{DC}$ for three AC field magnitudes ($E_{AC}$ =921.5,445.3,104μV/cm). SNR peaks between $E_{DC}$ =16mV/cm and 20mV/cm, with an average improvement of 6.05dB using an optimized DC field. The added DC Stark field alters the modulation sensitivity condition, enhancing or suppressing the FM sidebands as it changes how effectively the AC Stark modulation couples to the limit cycle oscillations. Additional data and analysis on power dependence as a function of DC Stark fields are provided in Supplementary Section 4.

### Discussion

Mode competition in nonequilibrium Rydberg gases unlocks new many-body phases[1-3], giving rise to collective oscillations or limit cycles that are not constrained by the energy levels of individual atoms[4,5]. Sustained limit cycle oscillations (OSC) emerge when driven by magnetic-fields (B-fields) at room temperature[6,16] due to the nonlinear energy shifts and competition between distinct Rydberg sublevels, giving rise to a long-range time crystalline order which manifests as a Rydberg dissipative time crystal (DTC). The sublevels are sensitive to DC or AC Stark E-fields, as they are shifted or modulated by these fields, thus shifting or modulating the OSC frequencies. The AC-Stark driven modulation of the OSC is shown as an effective technique to sense weak AC E-fields in the sub-kHz regime. With a modest setup, a sensitivity of ~7.8 μVcm$^{-1}$Hz$^{-1/2}$ for AC signals at 300Hz, and high-resolution detection to as low as sub-Hz is demonstrated. Even without system optimization, a ~8.7× improvement in sensitivity over state-of-art[17] Rydberg techniques is demonstrated. This approach facilitates the creation of ultra-compact (≪λ/10[6]) E-field detectors, with potential significance for applications in remote sensing, communications, navigation, and biomedical technology where ELF detectors are needed.

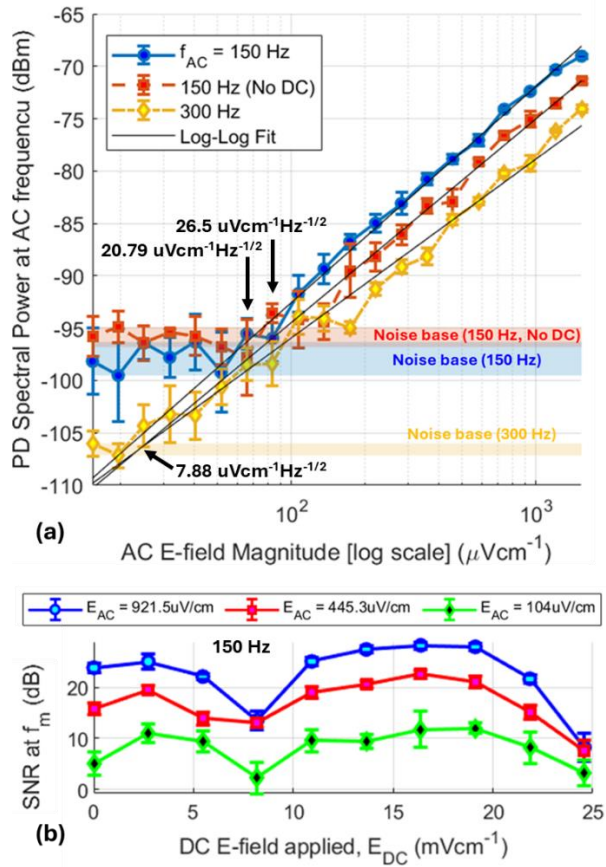

**Fig. 5:** Sensitivity analysis of AC E-field ($E_{AC}$) detection via modulation side-band magnitude detection and dependence on DC E-field ($E_{DC}$). (a) $E_{AC}$ is varied between 1.535mV/cm to 15.35μV/cm for $f_{AC}$ at 150Hz and 300Hz. At $f_{AC}$ =150Hz, measurements are conducted with and without a DC Stark field (blue line is with DC field, $E_{DC}$ =19.35mV/cm, which is close to the optimal point, see (b)). The DC field only slightly increases the signal-to-noise (SNR) ratio (improved sensitivity) at $f_{AC}$ =150Hz (20.79 μVcm$^{-1}$Hz$^{-1/2}$ with DC field to 26.5 μVcm$^{-1}$Hz$^{-1/2}$ without the DC field). At a higher modulation frequency of $f_{AC}$ =300Hz without the DC field, the noise-base reduction results in improved sensitivity to as high as 7.88 μVcm$^{-1}$Hz$^{-1/2}$. (b) The estimated SNR in dB at $f_{AC}$ =150Hz as a function of $E_{DC}$ is shown for three AC field measurements ($E_{AC}$ of ~921μV/cm, ~445μV/cm, ~104μV/cm). SNR is maximized between $E_{DC}$ =16mV/cm to 20mV/cm. Average SNR improvement with an optimized DC field was about 6.05dB. A resolution bandwidth of 10Hz is used in all measurements.

The AC Stark-driven modulation drives a frequency modulation (FM) of the OSC frequency and is easily extracted via standard demodulation techniques by detecting the tone at the peak of the modulation sideband. A DC field can selectively enhance the SNR by about 6.05dB (measured for AC signals at 150Hz), however that is not needed and not used for best-case sensitivity readings of ~7.8μVcm$^{-1}$Hz$^{-1/2}$ at 300Hz. Numerical simulations are presented using a V-type atomic structure for simplicity. Here AC Stark is used to modulate upper states of the V-type structure and are shown to closely reproduce measured time-frequency dynamics. The approach does not require optimization of the B-field magnitude to achieve any of the stated sensitivities. Furthermore, while the DC Stark field is shown to improve SNR slightly, it is not critically needed and can be avoided to give a simpler configuration compared to existing DC Stark-shift based techniques[17].

## Methods

### Mean-field dynamics with Stark modulation

The purpose of the model advanced here is to provide a qualitative description of the frequency modulation (FM) of the OSC frequency. This can be achieved using a simple V-type atomic structure[6,16] shown in Fig. 1d. Here the system is similar to the V-type system used to study sustained OSC in Rydberg time-crystals[6] but does not require the external on-resonant drive field to externally couple the upper states[16]. Instead, the atomic structure is modified to include Stark shifts and modulation of the upper states $|r\rangle$ and $|s\rangle$ (see Fig. 1d) via an external DC or AC E-field. The Hamiltonian in the rotating frame is:

$$\hat{H} = \frac{\Omega}{2}\sum_i \left(\hat{\sigma}_{gr}^i + \hat{\sigma}_{gs}^i + \text{H.c.}\right) - \sum_i \left(\Delta_r(t)\hat{n}_r^i + \Delta_s(t)\hat{n}_s^i\right) + \frac{1}{2}\sum_{i\neq j} V_{ij}\left(\hat{n}_r^i\hat{n}_r^j + 2\hat{n}_r^i\hat{n}_s^j + \hat{n}_s^i\hat{n}_s^j\right)$$

where $\hat{\sigma}_{\alpha\beta}^i = |\alpha^i\rangle\langle\beta^i|$ is the transition operator ($\alpha,\beta = g,r,s$), and $\hat{n}_\alpha^i = |\alpha^i\rangle\langle\alpha^i|$ ($\alpha = r,s$) denotes the local Rydberg density. $\Omega$ is the ground-to-excited state Rabi frequency. It is assumed that the interaction strengths, $V_{ij}$, is equal in magnitude to simplify development[6,16].

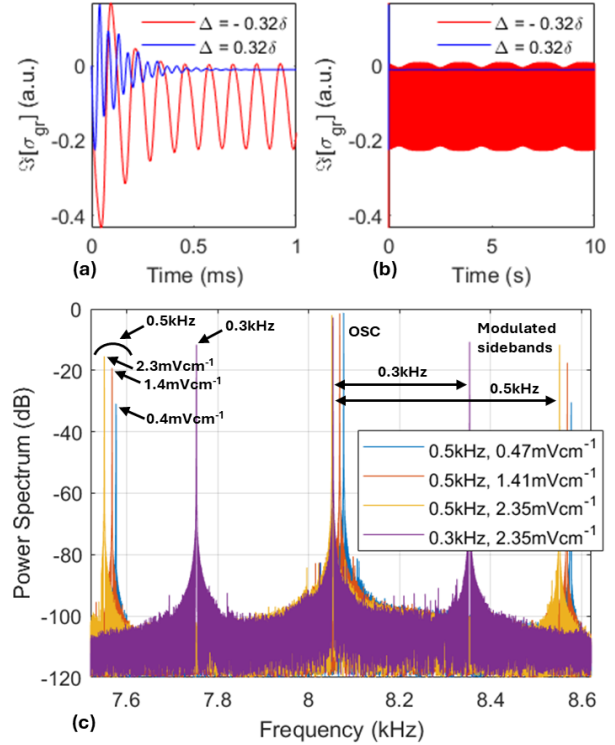

Fig. 6: Theoretical simulations of the imaginary part of $\sigma_{gr}$ based on the mean-field treatment under varying AC Stark modulations. (a) A generic V-type model is driven to exemplify the LCO's OSC transients in the <1ms regime (Ω/γ =3.29, Δ/δ =±0.32). Outside the OSC regime (Δ/δ =+0.32, blue) the $\Im[\sigma_{gr}]$ rapidly decays within ~0.5ms, whereas within the OSC regime (Δ/δ =-0.32, red), a self-sustained oscillation emerges. (b) When AC stark modulation with f$_{AC}$ =0.5Hz (α$_r$/δ=126.7 and α$_s$/δ=-32.8) is applied, a modulation of $\Im[\sigma_{gr}]$ is observed (red). (c) Applying an AC Stark modulation field (f$_{AC}$ =0.3kHz, 0.5kHz) results in sidebands at ±f$_{AC}$ relative to the OSC frequency when the spectrum of $\Im[\sigma_{gr}]$ is observed. Reduction in AC field magnitudes (calculated at f$_{AC}$ =0.5kHz) results in peak modulated signal intensity reduction at ±f$_{AC}$.

Since the AC Stark shifting is explicitly time-dependent due to the modulation, the detuning terms for $|r\rangle$ and $|s\rangle$ are written as $\Delta_r(t)$ and $\Delta_s(t)$. The system evolves on a characteristic timescale $T_{sys} \sim \max(\Omega^{-1}, \gamma^{-1})$, where $\gamma$ decay rate of the excited states, while the AC Stark modulation occurs over $T_{mod} \sim f_{mod}^{-1}$. The adiabatic approximation condition $T_{mod} \gg T_{sys}$ (or equivalently $f_{mod} \ll \max(\Omega, \gamma)$), ensures that the system can continuously adjust to the slowly varying detuning, justifying the quasi-static approximation. The system dynamics are governed by population equations, balancing excitation, decay, and coherence, shaped by interactions and Rabi oscillations. These have been explicitly derived in prior works[6,16], and as a result are not rederived here. In the mean-field approximation, inter-atom correlations are neglected, allowing factorization of higher-order moments: $\langle \hat{n}_r^i \hat{\sigma}_{gr}^j \rangle \approx \langle \hat{n}_r^i \rangle \langle \hat{\sigma}_{gr}^j \rangle$, $\langle \hat{n}_s^i \hat{\sigma}_{gs}^j \rangle \approx \langle \hat{n}_s^i \rangle \langle \hat{\sigma}_{gs}^j \rangle$. Assuming a uniform spatial distribution where $\langle \hat{n}_r^i \rangle = n_r$ and $\langle \hat{n}_s^i \rangle = n_s$, the nonlinear energy shift due to atomic interactions is given by: $E_{NL} = \chi(n_r + n_s)$, where $\chi$ captures the atomic interaction strength[6]. In the adiabatic limit, the mean-field population and coherence equations are:

$$\dot{n}_r \approx \frac{\Omega}{2}\Im(\sigma_{gr}) - \gamma n_r, \qquad \dot{n}_s \approx \frac{\Omega}{2}\Im(\sigma_{gs}) - \gamma n_s$$

$$\dot{\sigma}_{gr} \approx i\frac{\Omega}{2}(2n_r + n_s + \sigma_{sr} - 1)$$
$$\qquad + i\left(\Delta_r(t) - E_{NL} + i\frac{\gamma}{2}\right)\sigma_{gr}$$

$$\dot{\sigma}_{gs} \approx i\frac{\Omega}{2}(2n_s + n_r + \sigma_{rs} - 1)$$
$$\qquad + i\left(\Delta_s(t) - E_{NL} + i\frac{\gamma}{2}\right)\sigma_{gs}$$

$$\dot{\sigma}_{rs} \approx i\frac{\Omega}{2}(\sigma_{gs} - \sigma_{rg}) - i(\Delta_r(t) - \Delta_s(t) - i\gamma)\sigma_{rs},$$

$$\Delta_r(t) = \Delta_{r0} + \Delta_{AC,r}(t) \rightarrow \Delta_{AC,r}(t) = -\frac{1}{2}\alpha_r E^2(t)$$

$$\Delta_s(t) = \Delta_{s0} + \Delta_{AC,s}(t) \rightarrow \Delta_{AC,s}(t) = -\frac{1}{2}\alpha_s E^2(t)$$

with $\alpha_{r,s}$ being the polarizabilities of states $|r\rangle$ and $|s\rangle$. The AC Stark modulation $\Delta_r(t)$ and $\Delta_s(t)$ is observed to directly influence the phase evolution of the coherence terms $\sigma_{gr}$ and $\sigma_{gs}$, leading to a modulation of the limit cycle oscillations. Since these coherence terms drive population dynamics, their periodic modulation at the Stark shift frequency $f_{mod}$ induces corresponding variations in the oscillatory behavior of $n_r$ and $n_s$. This can be understood from the phase-dependent terms in the mean-field equations, where the coherence evolves as $\dot{\sigma}_{gr,s} \sim i(\Delta_{r,s}(t) - E_{NL})\sigma_{gr,s}$, showing that time-dependent detuning introduces a periodic phase modulation that propagates through the system, modifying the limit cycle oscillations.

Numerically solving the mean-field equations, $\Im(\sigma_{gr})$ is extracted which corresponds to probe transmission response from the ground to the intermediate state in experiments. Fig. 6 shows the mean-field simulations of for $\Im(\sigma_{gr})$ under different AC Stark modulations. Fig. 6a,b shows a characteristic transient solution for $\Im(\sigma_{gr})$ without OSC (blue), with OSC (red). In this example calculation, $\Omega/\gamma \sim 3.2$ and $\Delta/\delta \sim \pm 0.32$ (sign + is expected to show no OSC) is used to simply appreciate the nature of OSC transients (Fig. 6a) and modulation (Fig. 6b). Here $\Delta_{r0} = \Delta$ and $\Delta_{s0} = \Delta - \delta$. When outside the limit-cycle oscillation regime (blue), any initial oscillations decay rapidly, whereas in the OSC regime a self-sustained oscillation is formed (red). Using $\alpha_{r,s}$ =135.45, -35.15 MHz (computed by diagonalization of the Stark Hamiltonian within a defined basis of nearby states for $j$ =5/2, 3/2), and a $E_{AC}$ =3.3mV/cm AC field at 0.5Hz, Fig. 6b shows a modulation in the time-domain. The observed humps (red, Fig. 6b) arise from a slow phase modulation due to the time-dependent Stark shift, conceptually due to FM modulation but within the context of nonlinear atomic dynamics. Fig. 1c (bottom) showed a time-frequency analysis via a spectrogram, observing the deviation of the OSC frequency as a function of time. Fig. 6c studies the spectrum $\Im(\sigma_{gr})$ for sub-kHz AC Stark modulated fields. Applying an AC Stark modulation field ($f_{AC}$ =0.3 kHz, 0.5 kHz) introduces sidebands at ±$f_{AC}$ around the OSC frequency (which is at ~8.06kHz) in the spectrum of $\Im(\sigma_{gr})$. Decreasing the AC field magnitude (shown for $f_{AC}$ =0.5kHz) reduces the intensity of the modulated signal peaks at ±$f_{AC}$, demonstrating sensitivity of FM sideband magnitude to AC Stark field magnitude. In Fig. 6c, a DC field of 6.7mV/cm is used to reduce shifts in OSC frequencies.

**Experimental setup and approach**

The probe laser drives $6S_{1/2}$ F=4 to $6P_{3/2}$ F'=4 transition, with its frequency locked to the hyperfine structure using saturation absorption spectroscopy. The probe wavelength is $\lambda_p \sim$852.35672nm, with a linewidth <58kHz and detuning $\Delta_p/2\pi$ =22MHz. It has a Rabi frequency of $\Omega_p/2\pi$ =13.97MHz and a 1.1mm (1/e²) beam diameter. A half-wave plate (λ/2) sets the polarization before a Calcite beam displacer, which generates two orthogonally polarized probe beams with a 2.8mm offset. A quarter-wave plate (λ/4) ensures slight elliptical polarization, optimized to maximize signal-to-noise for the OSC and AC signals observed via probe balanced-amplified-photodetection (BAP) on a spectrum analyzer (SA). The beams pass through a vapor cell, centered between an embedded parallel-plate capacitor (PPC). The BAP output is split (-3dB RF splitter) between an oscilloscope and SA, with a 20kHz low-pass filter and a DC-block filter applied to suppress higher-order modes that could distort transient OSC signals, or DC offsets that could overload the SA.

The coupler laser, tuned to the $76D_{5/2}$ state, is frequency locked using a cylindrical cavity (50mm diameter × 100mm length, finesse $F_c \sim$15k placed in a vacuum housing with internal thermal shielding. The Pound-Drever-Hall (PDH) locking technique, combined with electronic sideband (ESB) modulation, stabilizes the coupler laser wavelength ($\lambda_c \sim$508.80582nm) and narrows its linewidth to <100Hz. The coupler detuning is $\Delta_c/2\pi$ =30MHz, with a Rabi frequency $\Omega_c/2\pi$ =0.85MHz and 1.3mm beam diameter. Both probe and coupler lasers are commercial external cavity diode lasers (ECDLs), with the ~509nm coupler generated via second harmonic generation (SHG).

The PPC embedded within the Cs vapor cell has a dimension of 18 × 45mm, separated by 15mm. Impedance transfer functions were measured with an impedance analyzer, and a fitted circuit model was used to calculate the relationship between voltage applied to the PPC and the resulting E-field magnitude[16]. The system was driven under a constant 4G B-field to develop sustained limit-cycle oscillations (OSC) between about 10.49-11kHz. This B-field was generated by a 30cm Helmholtz coil, spaced 15cm apart, driven by a DC supply.

Additional details of experimental systems and setup are presented in Supplementary Section 2.

### Data availability

The following are available as source data: 1) The LCO spectral data at n = 76 and B = 4G under different DC Stark-field (no modulation in Stark field) magnitudes (EDC up to 4.6mV/cm) in Fig. 2a; 2) Measurements of probe spectrum with OSC under different AC Stark modulation frequencies $f_{AC}$ in Fig. 4; 3) Sensitivity analysis datasets of AC E-field ($E_{AC}$) detection via modulation side-band magnitude when $E_{AC}$ is varied between 1.535mV/cm to 15.35μV/cm for $f_{AC}$ at 150Hz and 300Hz in Fig. 5a. All other data are available upon reasonable request.

**Figure Legends:**

**Fig. 1: (a) Experimental setup, simulated Stark modulations, energy diagram, and spectrogram measurements. (a)** Probe and coupler laser light is counter-propagated in a room-temperature vapor cell containing Cesium atoms. A parallel plate capacitor is embedded in the cell to drive low-frequency electric fields in the sub-kHz regime, and a DC magnetic field (B-field) is used to drive mode-competition at the excited Rydberg states to give limit cycle oscillations (OSC). A reference probe light is used for classical noise reduction. **(b)** Time-domain simulations of mode-competition between two excited states using a V-type atomic structure, driven additionally by a Stark modulation field at 0.5Hz (2 second period) is shown within (red) and outside (blue) of the OSC regime. In the OSC regime (red), the field modulates the energy levels of the two excited states and results in a modulation of the OSC frequency in the form of a frequency modulation. (c) Measurements and simulations of OSC at room-temperature is shown for a Stark field at 0.5Hz, 3.07mV/cm in a spectrogram. The low frequency of the Stark field (0.5Hz) slowly modulates the OSC frequency and is seen in the spectrogram as an approximately sinusoidal variation at 0.5Hz. Simulation based on a V-type Stark modulated atom system gives the same behavior. (d) The energy diagram in the ladder configuration used for experiments are composed of a probe laser driving to the first excited state at $6P_{3/2}$ and is detuned by $\Delta_p/2\pi$ = 22MHz, a coupler driving to $76D_{5/2}$ and is detuned by $\Delta_c/2\pi$ = 30MH, and a B-field of 4G is used. The Stark field modulates time-dependent shifts in the excited states that perturb the energy differences and OSC frequency. Simulations use a V-type model (gray line).

Fig. 2: (a) Limit-cycle oscillations (LCO) at n = 76 and B = 4G under different DC Stark-field (no modulation in Stark field) magnitudes ($E_{DC}$ up to 4.6mV/cm). The OSC frequency shifts gradually from 10.49kHz to 10.73kHz due to DC Stark shifting of the Rydberg states where mode-competition, LCO, and OSC arises. (b) A strong AC Stark modulation field ($E_{AC}$) is turned on with $f_{AC}$ =0.5Hz for 9.56s and then turned off. Stark-driven modulation of the OSC is seen at 0.5Hz. When $E_{AC}$ is turned off, the OSC frequency reverts to steady state at ~10.8kHz and is no longer AC modulated.

Fig. 3: Spectrograms and analysis of probe transmission signal to study OSC frequency deviation as a function of AC field magnitude. (a) AC Stark modulation frequency $f_{AC}$ =0.5Hz and $E_{AC}$ =3.07mV/cm. (b) $E_{AC}$ =1.53mV/cm. (c) $E_{AC}$ =0.31mV/cm. (d) Extracted peak points of each spectrogram dataset is plotted for each, showing gradual reduction frequency deviation. (e) Estimation of modulation depth, β, as a function of $E_{AC}$. Modulation depth is determined as the ratio of peak deviation to $f_{AC}$ (Δf/ $f_{AC}$) in percentages. A linear fit (black line), linear fit equation, and error bar is given.

Fig. 4: Measurements of probe spectrum with OSC under different and higher AC stark modulation frequencies $f_{AC}$, and with and without a DC offset at $f_{AC}$ =150Hz (blue vs red) to show differences in modulation peak. The spectral measurements (as opposed to spectrogram) show frequency modulation (FM) sidebands at $\pm f_{AC}$ on either side of the OSC frequency. To simplify comparisons, each are plotted as a function of (horizontal axis) $\Delta f_{OSC}$ (OSC shifted and centered at 0Hz). $E_{AC}$ =1.54mV/cm in all cases. At higher modulation frequencies (e.g.: >300Hz), peak at $f_{AC}$ gradually reduces in intensity, however the noise base is also reduced from being further away from the skirts of the LCO's OSC. The DC field of $E_{DC}$ =19.35mV/cm slightly increases the modulation sideband peak at $f_{AC}$ =150Hz.

Fig. 5: Sensitivity analysis of AC E-field ($E_{AC}$) detection via modulation side-band magnitude sensing and dependence on DC E-field ($E_{DC}$). (a) $E_{AC}$ is varied between 1.535mV/cm to 15.35μV/cm for $f_{AC}$ at 150Hz and 300Hz. At $f_{AC}$ =150Hz, measurements are conducted with and without a DC Stark field (blue line is with DC field, $E_{DC}$ =19.35mV/cm, which is close to the optimal point, see (b)). The DC field only slightly increases the signal-to-noise (SNR) ratio at $f_{AC}$ =150Hz (20.79 μVcm$^{-1}$Hz$^{-1/2}$ without DC field to 26.5 μVcm$^{-1}$Hz$^{-1/2}$ with the DC field). At a higher modulation frequency of $f_{AC}$ =300Hz without the DC field, the noise-base reduction results in improved sensitivity to as high as 7.88 μVcm$^{-1}$Hz$^{-1/2}$. (b) The estimated SNR in dB at $f_{AC}$ =150Hz as a function of $E_{DC}$ is shown for three AC field measurements ($E_{AC}$ of ~921μV/cm, ~445μV/cm, ~104μV/cm). SNR is maximized between $E_{DC}$ =16mV/cm to 20mV/cm. Average SNR improvement with an optimized DC field was about 6.05dB. A resolution bandwidth of 10Hz is used in all measurements.

Fig. 6: Theoretical simulations of the imaginary part of $\sigma_{gr}$ based on the mean-field treatment under varying AC Stark modulations. (a) A generic V-type model is driven to exemplify the LCO's OSC transients in the <1ms regime (Ω/γ =3.29, Δ/δ =±0.32). Outside the OSC regime (Δ/δ =+0.32, blue) the $\Im[\sigma_{gr}]$ rapidly decays within ~0.5ms, whereas within the OSC regime (Δ/δ =-0.32, red), a self-sustained oscillation emerges. (b) When AC stark modulation with $f_{AC}$ =0.5Hz ($\alpha_r/\delta$=126.7 and $\alpha_s/\delta$=-32.8) is applied, a modulation of $\Im[\sigma_{gr}]$ is observed (red). (c) Applying an AC Stark modulation field ($f_{AC}$ =0.3kHz, 0.5kHz) results in sidebands at $\pm f_{AC}$ relative to the OSC frequency when the spectrum of $\Im[\sigma_{gr}]$ is observed. Reduction in AC field magnitudes (calculated at $f_{AC}$ =0.5kHz) results in peak modulated signal intensity reduction at $\pm f_{AC}$.

### Acknowledgements

The author would like to acknowledge discussions with P. Mao and D. Willey at JPL (Jet Propulsion Laboratory, California Institute of Technology), A. Artusio-Glimpse, N. Prajapati, C. Holloway, and M. Simons at NIST (National Institute of Standards and Technology), and K. Cox, D. Meyer, and P. Kunz at ARL (Army Research Laboratory) as part of the NASA Instrument Incubator Program on Rydberg Radars. The research was carried out at the Jet Propulsion Laboratory, California Institute of Technology, under a contract with the National Aeronautics and Space Administration (80NM0018D0004), through the Instrument Incubator Program's (IIP) Instrument Concept Development (Task Order 80NM0022F0020).

### Author contributions

D.A conceived of the experiment and study reported, configured the atomic systems to include lasers and locking systems, and collected and processed all data reported in the text and figures. D.A. also developed all modeling, theoretical derivations and numerical simulations used or reported.

### Additional information

The author declares no competing interest.

March 10, 2025

## Supplementary information
## Supplementary Section 1
### Key features of a dissipative time crystal (DTC)

A dissipative time crystal (DTC) is characterized by four key criteria: (1) Spontaneous time-translation symmetry breaking (TTSB), where oscillations emerge intrinsically rather than being externally driven[7,24]; (2) Long-range temporal order, where persistent oscillations over extended timescales emerges and is sustained[25]; (3) Robustness against perturbations, ensuring oscillations persist despite some environmental noise or injected signals near the oscillation frequency, and parameter fluctuations[26]; and (4) Many-body interactions driving the oscillatory phase, distinguishing the system from a trivial few-body limit cycle[27]. These criteria have been experimentally demonstrated in Rydberg atoms: (i) Spontaneous emergence of oscillations is observed via sustained limit cycles in a Rydberg gas[4-6,16], confirming TTSB; (ii) Long-range order is verified by non-decaying autocorrelation functions[4,6,16], showing persistent phase coherence; (iii) Robustness to perturbations is confirmed by the system's resilience to external noise[6], signals in proximity to oscillation frequencies[16], and parameter variations such as B-fields that drive the sublevels[16]; and (iv) Many-body interactions are established through interaction-induced detuning shifts[6,16] and nonlinear competition between Rydberg states[4,6]. These results confirm the realization of a room-temperature dissipative time crystal (DTC) in a driven-dissipative Rydberg ensemble.

## Supplementary Section 2
### Additional details of experimental system and setup

The optical setup (see Extended Data Fig. 1) includes a frequency-locked Toptica DL Pro probe laser (~852nm) driving $6S_{1/2}$ F=4 to $6P_{3/2}$ F'=4. A half-wave plate sets the polarization, and a Calcite beam displacer splits the beam into two orthogonally polarized components, separated by 2.8mm. The probe light is adjusted to a slightly elliptical polarization using a quarter-wave plate (QWP) (see Extended Data Fig. 1), optimized for signal-to-noise ratio (SNR) as discussed in the main text. After passing through the vapor cell, the beams are detected by a Thorlabs PDB250A2 balanced detector. The resulting RF signal is filtered (DC-block and 20kHz low-pass filtered) before being analyzed on a spectrum analyzer (Keysight MXA N9021). The ~508nm coupling laser light is produced by second harmonic generation (SHG) of a ~1016nm light using a Toptica TA-MSHG PRO and drives the $6P_{3/2}$ F'=4 to $76D_{5/2}$ transition. The TA-MSHG PRO achieves second harmonic generation (SHG) using a nonlinear crystal inside a resonant enhancement cavity. The system first amplifies the fundamental ~1016 nm light using a tapered amplifier, then frequency doubles it to ~508 nm via a phase-matched nonlinear crystal. A HF-ANGSTROM WS/7 wavemeter is used to verify wavelengths of both probe and coupler lasers after they are frequency-locked with saturation absorption spectroscopy (SAS) or an optical cavity. The optical cavity used was a custom design by SLS (Stable Laser Systems). The cavity was placed inside a vacuum housing with multiple internal shields to ensure temperature control and was used to lock the coupling laser wavelength.

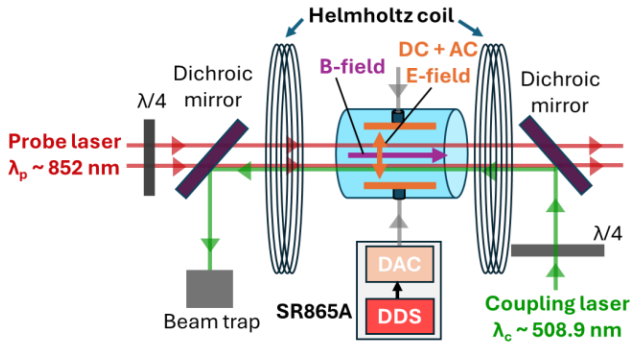

**Extended Data Fig. 1: System setup for experiments.** A Helmholtz coil is used to with 30cm diameter to generate uniform B-field at 4G with a separation of 15cm between coils. Vapor cell length was 56.6mm and diameter was 35mm. Distance between embedded plates for E-field generation is 15mm. The probe laser light drives the $6S_{1/2}$ F=4 to $6P_{3/2}$ F'=4. The coupler laser light drives the $6P_{3/2}$ F'=4 to $76D_{5/2}$. The probe (~852 nm) laser (DL Pro Toptica) is locked using a standard saturation absorption spectroscopy (SAS) lock. The coupler laser (TA-MSHG PRO Toptica) is locked to a medium finesse (~15k) optical cavity. An SR865A digital lock-in amplifier is used for signal generation (not detection) of both DC and AC signals.

The vapor cell is 56.6 mm long with a pair of embedded parallel stainless-steel electrodes (parallel-plate capacitor, PPC) measuring 18 × 45mm and separated by 15mm (Pyrex with no buffer gas). An SR865A digital lock-in amplifier is used for signal generation (not detection) of both DC and AC signals. It was chosen for precise DC and AC voltage generation due to its low-noise output, high-resolution digital control, and built-in signal synthesis, allowing fine-tuned voltage adjustments. The Helmholtz coil for B-field generation (3B Scientific 300 mm) consists of two identical coils, each with 320 windings, spaced apart to generate a uniform magnetic field in the central region.

Spectrogram measurement data was generated directly by the Keysight MXA N9021 with an acquisition time of ~0.08s and resolution bandwidth of ~9.69Hz. All spectral data and sensitivity analysis was conducted using the Keysight MXA N9021 in spectrum mode (not spectrogram).

### Supplementary Section 3
### Deviation of sensitivity from fitted curves

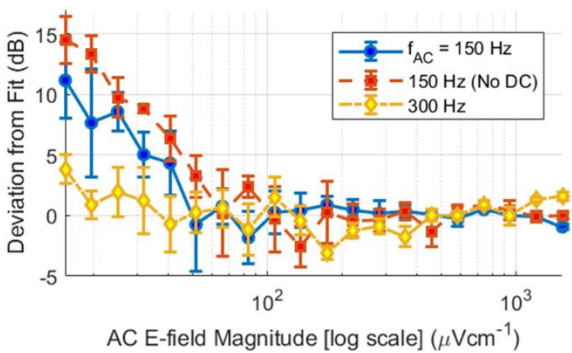

**Extended Data Fig. 2:** Deviation from best case fit lines in Fig. 5a of the main text. RMS deviation from fit up to the estimated sensitivity was found to be approximately 0.68dB, 1.09dB, and 1.29dB for $f_{AC}$ =150Hz with $E_{DC}$ =19.35mV/cm, $f_{AC}$ =150Hz without DC, and $f_{AC}$ =300Hz without DC, respectively.

In Fig. 5a (main text), the AC Stark E-field was varied from 1.535mV/cm to 15.35μV/cm at 150Hz and 300Hz. Sensitivity at 150Hz improved by ~27% with a DC Stark field (19.35mV/cm), reaching 20.79μVcm$^{-1}$Hz$^{-1/2}$, while at 300Hz, noise reduction improved sensitivity to 7.88 μVcm$^{-1}$Hz$^{-1/2}$. Log-log curve fits (black lines in Fig. 5a main text) are studied from the perspective of deviation analysis in Extended Data Fig. 2. The curve fits are obtained by performing a log-log linear regression on the measured spectral power of the modulation sideband peak versus driven AC field magnitude, excluding a predefined number of weakest signal points to avoid noise floor effects. The resulting fit function is subtracted from the measured data to compute the deviation in Extended Data Fig. 2, and error analysis is performed to assess the quality of the fit. The RMS (root-mean-squared) deviation from the log-log fit, up to the estimated sensitivity limit, was measured as ~0.68dB for $f_{AC}$ =150Hz with $E_{DC}$=19.35mV/cm, ~1.09dB for $f_{AC}$ =150Hz without DC, and ~1.29dB for $f_{AC}$ =300Hz without DC. Here only points above the noise floor is used, which is different for each case. As shown in Extended Data Fig. 2, the deviation is highest at lower AC field magnitudes, particularly for the 150Hz cases, and stabilizes at larger fields, indicating reduced relative error in stronger signal regimes as is expected.

### Supplementary Section 4
### Power and SNR as a function of DC Stark field

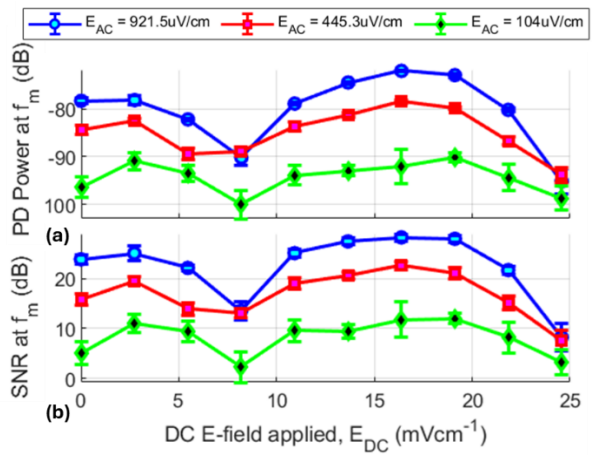

**Extended Data Fig. 3:** The measured power (a) and estimated SNR (b) in dB at $f_{AC}$ =150Hz as a function of $E_{DC}$ is shown for three AC Stark E-field magnitudes ($E_{AC}$ =921μV/cm, 445μV/cm, 104μV/cm). As $f_{AC}$ =150Hz is fixed, the noise-base is not to significantly varying between measurements, and as a result, both power (a) and SNR (b) is maximized between $E_{DC}$ =16mV/cm to 20mV/cm. A resolution bandwidth of 10Hz is used in all measurements.

In Fig. 5b (main text), the signal-to-noise ratio (SNR) dependence on $E_{DC}$ (up to 24.57mV/cm) was analyzed for $f_{AC}$ =150Hz. For $E_{AC}$ =921.5, 445.3, and 104 μV/cm, SNR peaks between 16–20 mV/cm, showing an average 6.05dB improvement with optimized DC fields. While SNR is ultimately of interest, it is useful to investigate both power of the modulated sideband and SNR. Extended Data Fig. 3 gives the measured photodetector (PD) power of the FM modulated sideband (a) and estimated signal-to-noise ratio (SNR) (b) as functions of the applied DC Stark field ($E_{DC}$ at $f_{AC}$ =150Hz for the three AC field magnitudes (for $E_{AC}$ =921.5, 445.3, and 104 μV/cm). Extended Data Fig. 3b is identical to what is presented in Fig. 5b (main text), repeated here for convenience. Since the noise floor remains relatively stable with a fixed $f_{AC}$ in this analysis, both PD power and SNR maximize in the range of $E_{DC}$ =16-20mV/cm, indicating an optimal DC field for enhancing signal strength. The shape of the curves (both PD power and SNR) observes similar trends, showing that as power increases (decreases), SNR increases (decreases), thus suggesting noise floor remains relatively stable. The DC Stark shift shifts the energy levels relative to the AC field modulation. If the DC shift moves the energy levels into a regime where the AC-induced modulation is stronger, the FM sidebands will be enhanced. If the DC shift moves the levels away from this condition, the modulation depth decreases, and the FM sidebands are weaker. This behavior leads to a trend in the strength of the FM sideband as a function of $E_{DC}$, because at certain values of $E_{DC}$, the modulation depth is maximized, while at other values, it is minimized. A 10Hz resolution bandwidth is used in all measurements to maintain consistency.

**Additional References for Supplementary Sections**